\documentstyle[preprint,aps,epsfig]{revtex}
\draft
\begin{document}
\preprint{\bf\it Phys. Rev. C (2001) in press.}
\title{Proton Differential Elliptic Flow and the Isospin-Dependence of 
the Nuclear Equation of State}
\bigskip
\author{\bf Bao-An Li\footnote{email: Bali@astate.edu},
Andrew T. Sustich\footnote{email: Sustich@astate.edu} 
and Bin Zhang\footnote{email: Bzhang@astate.edu}}
\address{Department of Chemistry and Physics\\
P.O. Box 419, Arkansas State University\\
State University, Arkansas 72467-0419, USA}
\maketitle

\begin{quote}
Within an isospin-dependent transport model for nuclear reactions involving
neutron-rich nuclei, we study the first-order direct transverse flow of protons
and their second-order differential elliptic flow as a function of transverse
momentum. It is found that the differential elliptic flow of mid-rapidity 
protons, especially at high transverse momenta, is much more sensitive to the isospin 
dependence of the nuclear equation of state than the direct flow. Origins of 
these different sensitivities and their implications to the experimental 
determination of the isospin dependence of the nuclear equation of state by 
using neutron-rich heavy-ion collisions at intermediate energies are discussed\\ 
{\bf PACS} numbers: 25.70.-z, 25.75.Ld., 24.10.Lx
\end{quote}

\newpage
The isospin-dependence of the nuclear equation of state {\rm (EOS)} is one of
the most important but very poorly known property of neutron-rich 
matter\cite{ibook}. Its determination in laboratory-controlled experiments
has profound implications to the study of the structure and evolution of 
many astrophysical objects\cite{lat00}. Nuclear reactions induced by stable neutron-rich
nuclei and/or radioactive beams provide a means to extract useful information
about the isospin-dependence of the nuclear {\rm EOS} and to explore novel 
phenomena in nuclear matter at extreme isospin asymmetries. A number of 
dedicated experiments to study the isospin dependence of the nuclear 
{\rm EOS} have been performed/planned at several available radioactive 
beam facilities and the future Rare Isotope Accelerator {\rm (RIA)}\cite{ria}. 
For these experiments to be fruitful it is important to first understand well 
the role of the isospin degree of freedom in nuclear reaction dynamics. 
Moreover, theoretical predictions on the sensitivity of experimental 
observables to the isospin dependence of the nuclear {\rm EOS} are useful 
for explaining available data and planning new experiments. Recently, 
several useful observables have been 
identified in neutron-rich heavy-ion collisions at intermediate energies. These 
include the isospin fractionation\cite{li97,li98,xu00,yel01,udo01,ditoro01,li01}, 
neutron to proton ratio of heavy residues\cite{tan01} or projectile-like 
fragments\cite{yel01,sjy00} and the neutron-proton differential flow\cite{li00}.
Most of these observables make use of the relative multiplicities and 
kinetic energies of mirror nuclei. Information about the isospin dependence 
of the nuclear {\rm EOS} can also be obtained from studying free nucleons as 
neutrons and protons have the opposite symmetry potentials in nuclear medium. 
In a recent work, one of the present authors has shown that the neutron-proton 
differential flow is a rather useful probe of the isospin dependence of the 
nuclear {\rm EOS}. The method utilizes constructively both the isospin 
fractionation and the nuclear collective flow as well as their sensitivities
to the nuclear {\rm EOS}. It, however, requires measuring the neutron and proton 
collective flow simultaneously. Experimentally, it is easier to identify 
charged particles and measure their momenta accurately, thus observables 
using charged particles only are more useful. In this work, within an 
isospin-dependent transport model for nuclear reactions involving 
neutron-rich nuclei, we explore the sensitivity of proton collective flow 
to the isospin dependence of the nuclear {\rm EOS} in heavy-ion collisions 
at intermediate energies. We examine both the first-order transverse flow and
the second-order differential elliptic flow as functions of transverse momentum. 
It is found that the differential elliptic flow of midrapidity protons, 
especially at high transverse momenta, is very sensitive to the isospin 
dependence of the nuclear {\rm EOS}. This sensitivity is much stronger 
than that found in the first-order transverse collective flow around the 
projectile and target rapidity.\  
  
At present, nuclear many-body theories predict vastly different isospin
dependence of the nuclear {\rm EOS} depending on both the calculation 
techniques and the bare two-body and/or three-body interactions 
employed, see e.g., \cite{wir88,akm97,brown00,hor00}. Various theoretical 
studies (e.g., \cite{bom91,hub93}) have shown 
that the energy per nucleon $e(\rho,\delta)$ in nuclear matter 
of density $\rho$ and isospin asymmetry parameter $\delta$ defined as
\begin{equation}
\delta\equiv(\rho_n-\rho_p)/(\rho_n+\rho_p)
\end{equation}
can be approximated very well by a parabolic function
\begin{equation}\label{aeos}
e(\rho,\delta)= e(\rho,0)+S(\rho)\cdot\delta^2.
\end{equation}
In the above $e(\rho,0)$ is the {\rm EOS} of isospin symmetric nuclear matter 
and $S(\rho)$ is the symmetry energy at density $\rho$. The {\rm EOS}
should also be momentum-dependent, however, for this exploratory study we shall negelect this dependence. 
We shall use for isospin-symmetric nuclear matter a stiff {\rm EOS} with $K_0=380$ MeV which can reproduce the transverse 
flow data equally well as a momentum-dependent soft
{\rm EOS} with $K_0=210$ MeV\cite{pan93,zhang93}. This choice will not affect our
conclusions since the elliptic flow has been shown to be insensitive to the
momentum-dependence of the {\rm EOS} in heavy-ion collisions at beam energies 
below about 1.5 GeV/nucleon if $K_0=380$ MeV is used\cite{pawel98}.  
The form of the 
symmetry energy as a function of density is rather strongly model 
dependent\cite{bom01}. Very divergent predictions on the isospin 
dependence of the nuclear {\rm EOS} by various many-body theories 
have lead to vastly different forms for the $S(\rho)$. We adopt here 
a parameterization used by Heiselberg and Hjorth-Jensen in their studies 
on neutron stars\cite{hei00}
\begin{equation}\label{srho2}
S(\rho)=S_0(\rho_0)\cdot u^{\gamma},
\end{equation}
where $u\equiv \rho/\rho_0$ is the reduced density and $S_0(\rho_0)$ is 
the symmetry energy at normal nuclear matter density $\rho_0$. The value of
$S_0(\rho_0)$ is known to be in the range of about 27-36 {\rm MeV} from 
analyzing atomic masses\cite{farine,mass,pear}. By fitting the result of 
variational many-body calculations by Akmal et al\cite{akm97}, Heiselberg and Hjorth-Jensen found the values of 
$S(\rho_0)=$32 MeV and $\gamma=0.6$. However, as shown by many other authors 
previously\cite{bom01} and more recently by Brown\cite{brown00} using other 
approaches, the extracted value of $\gamma$ varies widely, even its sign is 
undetermined. Therefore,in this work we take the $\gamma$ as a free parameter and study 
its influence on the experimental observables in neutron-rich heavy-ion 
collisions. Further, we take a constant value of $30$ MeV for $S_0(\rho_0)$ 
in the following study.

It is well known that the symmetry energy has a kinetic and a potential
contribution
\begin{equation}\label{srho}
 S(\rho)=\frac{3}{5}e_{\rm F}^0u^{\frac{2}{3}}(2^{\frac{2}{3}}-1)+V_2
\end{equation}
where $e_{F}^0$ is the Fermi energy in symmetric nuclear matter at normal density and $V_2$ is the 
potential contribution. Within our transport model, the kinetic part of the symmetry energy is 
simulated by using different Fermi momenta for neutrons and protons according to the local 
Thomas-Fermi approximation. While the potential part is taken into account by using the 
symmetry potential. Corresponding to Eq. \ref{srho} the symmetry potential energy density is
\begin{equation}
W_{asy}=V_2\rho\delta^{2},
\end{equation}
and the single-particle symmetry potential
$V_{\rm asy}^{q}$ can be obtained from
\begin{equation}\label{vasy}
V_{\rm asy}^{q}=\frac{\partial W_{asy}}{\partial \rho_{q}}
=(S_0(\gamma-1)u^{\gamma}
+4.2u^{\frac{2}{3}})\delta^{2}\pm(S_0u^{\gamma}-12.7u^{2/3})\delta
\end{equation}
where ``+" and ``-" are for $q=neutron$ and $q=proton$, respectively.
For small isospin asymmetries and densities near $\rho_0$ the above 
symmetry potentials reduce to the well-known Lane potential which
varies linearly with $\delta$\cite{lane}. Shown in Fig.\ 1 are the symmetry 
potentials at $\delta$=0.2 with the $\gamma$ parameter of 0.5, 1.0 and 2.0 for 
neutrons (upper branch) and protons (lower branch), respectively. 
Generally, the repulsive/attractive symmetry potential for neutrons/protons
increases with density. 
The isospin dependence of the nuclear {\rm EOS} can be characterized by the 
curvature of the symmetry energy at $\rho_0$\cite{bom91} 
\begin{equation}
K_{\rm sym}\equiv 9\rho_0^2\frac{\partial^2 S(\rho)}{\partial \rho^2}|_{\rho
=\rho_0}=9S_0(\rho_0)\cdot\gamma(\gamma-1).
\end{equation}
The slope and curvature of the symmetry potential 
depends strongly on the parameter $\gamma$. We consider in the following two 
cases with $\gamma=0.5$ and 2, the corresponding $K_{sym}$ parameter is $-68$ {\rm MeV} 
and $540$ {\rm MeV}, respectively. This choice of the parameter $\gamma$ 
allows us to explore a large uncertain range of the isospin dependence of the 
nuclear {\rm EOS}. 

Our study is based on the isospin-dependent Boltzmann-Uehling-Uhlenbeck (IBUU) 
transport model (e.g., \cite{li97,ls99}). In this model protons and 
neutrons are initialized in coordinate space according to their density 
distributions predicted by the relativistic mean-field (RMF) theory\cite{serot,ren}. In momentum 
space they are initialized in Fermi spheres with individual neutron or proton 
Fermi momentum dertermined by its local density
according to the Thomas-Fermi approximation. The isospin-dependent 
reaction dynamics is included through isospin-dependent nucleon-nucleon 
scatterings and Pauli blockings, the symmetry potential $V_{\rm asy}$ and the 
Coulomb potential $V_c^p$ for protons. A Skyrme-type parameterization is 
used for the isoscalar potential. For a review of the model, we refer the
reader to ref.\ \cite{li98}. However, a few comments relevant to the present study
are necessary here. We did not include in the EOS the curvature term which is 
important for reproducing surface properties of static nuclei and keeping them stable. 
This is because we can not find a universal coefficient of this term that can reproduce the 
neutron and proton density distribtions of all nuclei up to mass 200. 
Instead, we used the predictions by the  RMF theory to initialize
nucleons. It would be ideal if one can use the same model with the same interaction to first reproduce 
perfectly the initial neutron and proton density and momentum distributions, in particular, the neutron 
skins of heavy nuclei, then study the reactions between them. 
What we have done sofar is thus a compromise. Nevertheless, the best available knowledge 
about the inital neutron and proton distributions is used, and we found that the initial 
distributions for stable nuclei are kept stable upto about 150 fm/c with energetic 
particle emissions less than about 1 percent in our approach. 
Thus the distributions are sufficiently stable over a time period that
is long enough for studying the collective flow which is mainly generated in the
early stage of the reaction.

First, to understand the role of the symmetry potential in the reaction 
dynamics we have studied the central density as a function of time for 
the reaction of Sn+Sn. To the first order of approximation, there is 
essentially no difference in the evolution of the central density with 
$\gamma=$0.5 and 2 since the symmetry potential is rather small compared 
to the isoscalar nuclear potential. We have also investigated the relative roles of the
symmstry and Coulomb potentials in the reaction dynamics by turning on or off one 
of them. We found that the Coulomb potential dominates over the attractive symmetry 
potential for protons for the reaction considered here. However, without including 
the symmetry potential the Coulomb potential along will lead
artifically to more proton emissions than neutrons in the early stage of the reaction.
Moreover, stable nuclei can not be kept stable using only the Coulomb potential 
without the symmetry potential in the transport model. We thus either turn on or off 
simultaneously both the Coulomb and symmetry potential in our following studies. 
To evaluate the degree of isospin asymmetry obtained and its correlation with density 
we show in Fig.\ 2 a scatter plot of the local density $\rho(cell)/\rho_0$ and the isospin asymmetry 
$\delta(cell)$ of all non-vacant cells of 1 $fm^3$ volume each. The plot is 
made by using the parameter $\gamma$=0.5 for the Sn+Sn reaction at the time 
instant of 30 fm/c when the central density is close to its maximum. It is seen 
that the high density region centers around the isospin asymmetry of the reaction 
system of $\delta_0\approx 0.19$ as one would expect. Cells with lower 
densities extends in $\delta$ toward the neutron-rich side far away from 
$\delta_0$ due to both the neutron-skins of the colliding nuclei and the 
isospin fractionation during the reaction. The latter is an unequal 
partitioning of the neutron to proton ratio $N/Z$ of asymmetric nuclear matter 
between low and high density regions. It is energetically favorable for the
asymmetric nuclear matter to be separated into a neutron-rich low density 
phase and an isospin-symmetric high density one\cite{muller,liko,baran,shi,cat}. 
Since the symmetry potential is an increasing function of $\delta$, the symmetry 
potentials at densities less than $\rho_0$ but with larger isospin asymmetries 
are also appreciable as in the higher density region around the $\delta_0$.
Although this symmetry potential itself has almost no effect on the compression and
expansion of the central region, it does have a significant
effect on the emission of particles and their kinetic energy spectra. As seen 
from Fig.\ 1, at densities below about $\rho_0$, the attractive symmetry 
potential for protons is stronger with the parameter $\gamma=$0.5 than $\gamma=2.0$. 
As a result, more protons are expected to become unbound with $\gamma=$2. 
Shown in Fig.\ 3 are the unbound proton rapidity distributions at the freeze-out 
for the Sn+Sn reaction. The unbound protons are identified as those with local densities less than 
$\rho_0/8$ and the freeze-out time is found to be about 100 fm/c for the 
reaction considered here. Appreciable effects of the symmetry potential on 
the rapidity distribution of free protons are seen over the whole rapidity 
range. Significantly more particles are emitted with $\gamma=2.$, especially 
around the midrapidity. This observation is consistent with the total and 
differential isospin fractionation discussed in detail in 
refs. \cite{li97,li01,li00}.

We now turn to the exploration of the isospin dependence of the nuclear 
{\rm EOS} by studying both the first-order transverse collective flow and 
the second-order differential elliptic flow of unbound protons. Collective
flow has been well known for its ability to reveal the pressure gradient 
during nuclear reactions, see e.g., \cite{gupta,pawel,gary}. In the local rest 
frame of matter the collective flow velocity $\vec v$, energy density $e$ 
and pressure $p$ are related by the relativistic Euler equation\cite{pawel}
\begin{equation}
(e+p)\frac{\partial}{\partial t}\vec v=-\vec{\nabla}p.
\end{equation}
The pressure gradient $\vec{\nabla}p$ plays the role of a driving 
force for the acceleration of the collective flow velocity $\vec v$. 
For isospin asymmetric nuclear matter the pressure $p$ 
has a kinetic part $p_{kin}$, an isospin-independent interaction part 
$p_0$ and an isospin-dependent one $p_{asy}$, i.e., 
\begin{equation}
p=p_{{\rm kin}}+p_0+p_{{\rm asy}}.
\end{equation}
At temperature $T$ the kinetic contribution $p_{{\rm kin}}$ is 
\begin{equation}
p_{{\rm kin}}=T\rho\{1+\frac{1}{2}\sum_{n=1}^{\infty}
b_n(\frac{\lambda_T^{3}\rho}{4})^n\left[(1+\delta)^{1+n}
+(1-\delta)^{1+n}\right]\},
\end{equation}
where the $b_n's$ are the inversion coefficients given in 
ref. \cite{li98} and
\begin{equation}
\lambda_T=\left(\frac{2\pi \hbar^2}{mT}\right)^{1/2}
\end{equation}
is the thermal wavelength of nucleons with mass $m$. The contribution 
from the isospin-independent nuclear interaction $p_0$ is
\begin{equation}
p_0=\frac{1}{2}a\rho_0(\frac{\rho}{\rho_0})^2+\frac{b\sigma}
{1+\sigma}\rho_0(\frac{\rho}{\rho_0})^{\sigma+1},
\end{equation}
where $a=-123.6$ MeV, $b=70.4$ MeV and $\sigma=2$ corresponding to the stiff 
nuclear {\rm EOS} of isospin-symmetric nuclear matter. 
The asymmetry pressure $p_{{\rm asy}}$ due to the isospin-dependent 
potential is
\begin{equation}
p_{asy}=\delta^2\left[S_0(\rho_0)\cdot\rho_0\cdot\gamma u^{\gamma+1}-8.5 u^{5/3}\right].
\end{equation}
Shown in Fig.\ 4 is a comparison of the density derivatives of the 
three partial pressures. As an illustration we have selected the parameter set
of $\delta=0.2$, $T=5$ MeV, $\gamma$=0.5 and 2. In the whole density range 
reachable in heavy-ion collisions at intermediate energies, the gradient of the 
total pressure is overwhelmed by the contributions from the 
isospin-independent interaction and the kinetic pressure. The contribution from
the isospin-dependent interaction is relatively small. Nevertheless, it is still
useful to note that both the symmetry pressure and its density gradient 
increase with the $\gamma$ parameter, especially at high densities. 
We thus expect to see a stronger flow signal with a larger $\gamma$ parameter, 
but the effect should be rather small. This expectation is verified 
by performing the standard analysis of the transverse collective 
flow\cite{pawel85} 
\begin{equation}
< p_x/A >(y)=\frac{1}{A(y)}\sum_{i=1}^{A(y)} p_{ix}(y),
\end{equation}
where $A(y)$ is the number of particles at rapidity $y$ and $p_{ix}$ is $i^{th}$ 
particle's transverse momentum in the reaction plane. 
As shown in Fig.\ 5., a slightly stronger transverse flow signal is obtained
with $\gamma$=2 than $\gamma$=0.5 as expected. Given the difficulties and 
uncertainties in determining the reaction plane, the observed effect of the
isospin dependence of the nuclear {\rm EOS} is too small to be practically
very useful. To explore the isospin-dependent {\rm EOS}, one thus has to
look for more delicate observables, such as the neutron-proton differential
flow as proposed in ref. \cite{li00}. 

As a complementary approach which also has the advantage of not having 
to measure the collective flow for neutrons, we investigate the
proton differential elliptic flow at midrapidity as a function of transverse
momentum\cite{oll,art}
\begin{equation}
<v_2(p_{t})>=\frac{1}{N}\sum_{i=1}^{N}\frac{p_{ix}^2-p_{iy}^2}{p_{it}^2},
\end{equation}
where $N$ is the total number of unbound protons in the rapidity range
of $-0.5\leq (y/y_{beam})_{cms}\leq 0.5$. In this rapidity range 
the isospin fractionation is the strongest, i.e., the value of $N$ depends
most sensitively on the parameter $\gamma$, as shown in Fig.\ 3. The $p_{iy}$
is the ith particle's transverse momentum perpendicular to the reaction plane.   
The value of $v_2(p_t)$ is thus a differential measure of the second-order 
flow effect as a function of transverse momentum. The sign and magnitude
of $v_2$ reflects the result of a competition between the early 
``squeeze-out''($p_{iy}$ dominates) perpendicular to the reaction plane and 
the later in-plane transverse flow of nucleons. The differential 
elliptic flow, especially at high $p_t$, is expected to be more 
sensitive to the isospin dependence of the nuclear {\rm EOS} than the 
first-order transverse flow. This is because all three partial pressures 
lead approximately to a similar difference $\delta p_{xy}^2\equiv p_x^2-p_y^2$ 
although their respective contributions to the value of $<p_x>$ or $<p_y>$ 
is very different. Moreover, the early pressure created in the participant 
region is revealed more clearly by the value of $v_2(p_t)$ at high transverse 
momenta. This is due to the fact that high $p_t$ particles can only be 
produced through the most violent collisions in the early stage of the reaction. 
This fact is almost universal in heavy-ion collisions at all 
energies, see, e.g., \cite{bli94,pawel95,liz99,gyu}.
These particles can only retain their high transverse momenta by escaping from 
the reaction zone along the direction perpendicular to the reaction plane 
without suffering much rescatterings. 

Shown in Fig.\ 6 are the predicted proton differential elliptic 
flow as a function of transverse momentum $p_t$. The results presented here are
obtained by using 20,000 Sn+Sn events in each case. Below the Fermi 
momentum of about 0.3 GeV/c the elliptic flow increases with $p_t$ as 
predicted by the nuclear hydrodynamics\cite{hen99,heinz}. This is 
because the low $p_t$ particles have undergone a sufficiently large number of 
rescatterings to reach the hydrodynamical limit and to obtain an in-plane 
flow indicated by the positive value of $v_2(p_t)$. Particles with higher transverse 
momenta are mainly those emitted perpendicular to the reaction plane with 
negative values of $v_2$. These particles carry more undisturbed information 
about the initial high density phase of the reaction. This effect becomes stronger gradually 
with the increasing transverse momentum. Therefore the $v_2(p_t)$ value 
decreases with the increasing $p_t$ above the Fermi momentum. Indeed, it is 
seen that the elliptic flow at high $p_t$ is much more sensitive to the parameter 
$\gamma$ than the first-order transverse flow. Without the Coulomb repulsion 
the shadowing effect of the spectator nucleons lasts longer, it leads to a 
more stronger ``squeez-out'' and thus a more negative value of elliptic flow, 
especially at high $p_t$. As the pressure gradient increases by increasing 
the parameter $\gamma$, the in-plane flow becomes dominant over 
the ``squeeze-out'' in the whole range of transverse momentum. The value of 
$v_2(p_t)$ thus increases with the parameter $\gamma$, particular in the high 
$p_t$ region. The sensitivity observed here is much stronger than that 
observed in analyzing the first-order transverse flow. Although very high 
$p_t$ particles are rather rare, it is practically relatively easy for 
nucleons to obtain a transverse momentum upto 2 times the Fermi momentum 
in heavy-ion collisions at intermediate energies. In the range of 
$p_t\leq 0.5$ GeV/c, our results 
here indicate an unambiguously strong signal of the isospin dependence of the
nuclear {\rm EOS} far beyond the statistical error bars.   

In summary, within an isospin-dependent transport model for nuclear reactions 
involving neutron-rich nuclei, we analyzed the first-order direct flow and the
second-order differential elliptic flow for free protons in the reaction of
$^{124}Sn+^{124}Sn$ at a beam energy of 50 MeV/nucleon and an impact parameter 
of 5 fm. The emphasis was placed on searching for observable signals of the 
isospin dependence of the nuclear {\rm EOS}. We found that the differential 
elliptic flow, especially at high transverse of momenta, for midrapidity protons 
is much more sensitive to the isospin dependence of the nuclear {\rm EOS} than 
the first-order direct flow. This finding provides a useful guide for planning
future experiments to study the isospin dependence of the nuclear {\rm EOS} 
by using neutron-rich heavy-ion collisions at intermediate energies.

This work was supported in part by the National Science Foundation 
Grant No. PHY-0088934, Arkansas Science and Technology Authority Grant 
No. 00-B-14 and Grant No. 01-B-20.

\newpage
\begin{figure}[htp] 
\centering \epsfig{file=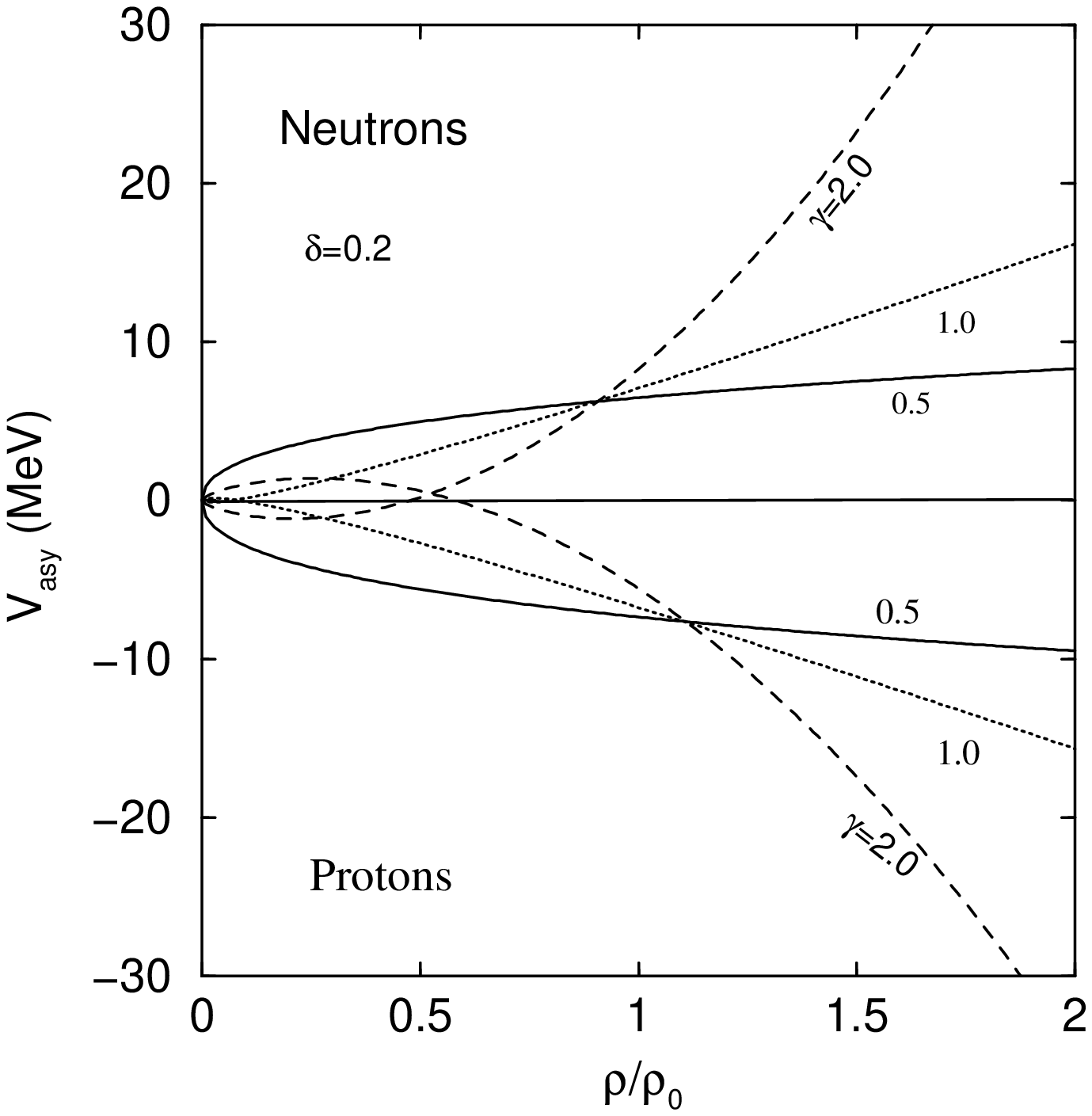,width=12cm,height=12cm,angle=-90} 
\caption{Symmetry potential for neutrons (upper branches) and protons (lower
branches) as a function of density for an isospin asymmetry of $\delta=0.2$
and the $\gamma$ parameter of 0.5, 1.0 and 2.0, respectively.} 
\label{fig1} 
\end{figure} 

\begin{figure}[htp] 
\centering \epsfig{file=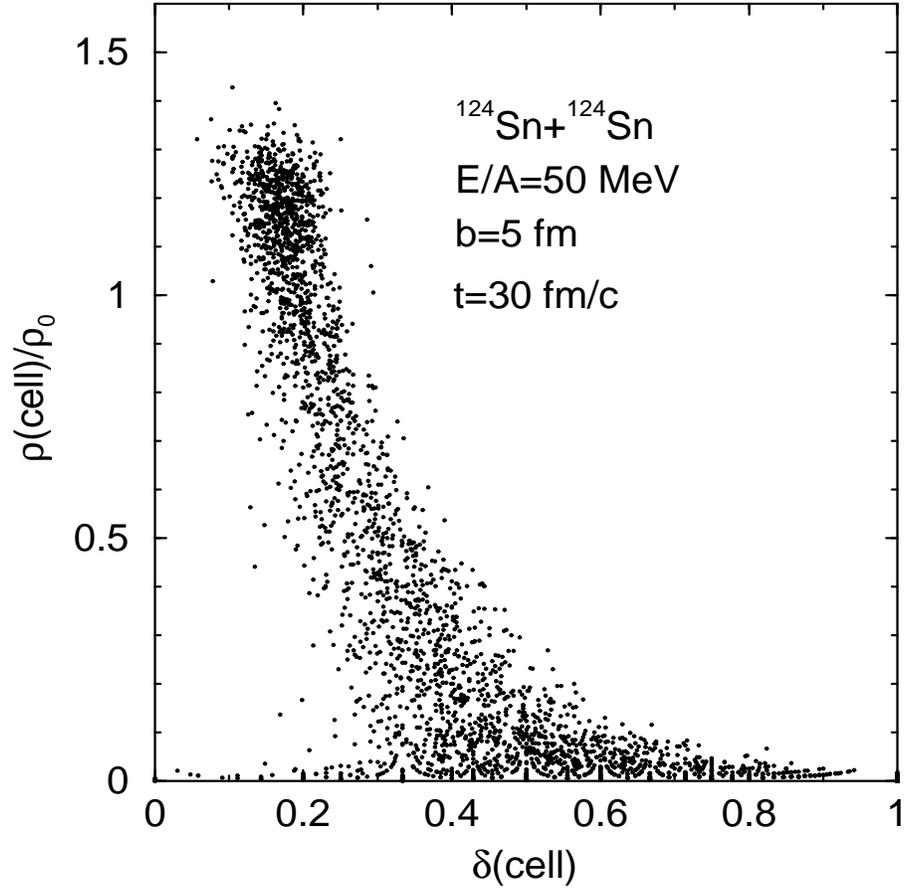,width=12cm,height=12cm,angle=-90} 
\caption{A scatter plot of the nucleon density of each cell versus its
isospin asymmetry $\delta$.}
\label{fig2}
\end{figure}

\begin{figure}[htp] 
\centering \epsfig{file=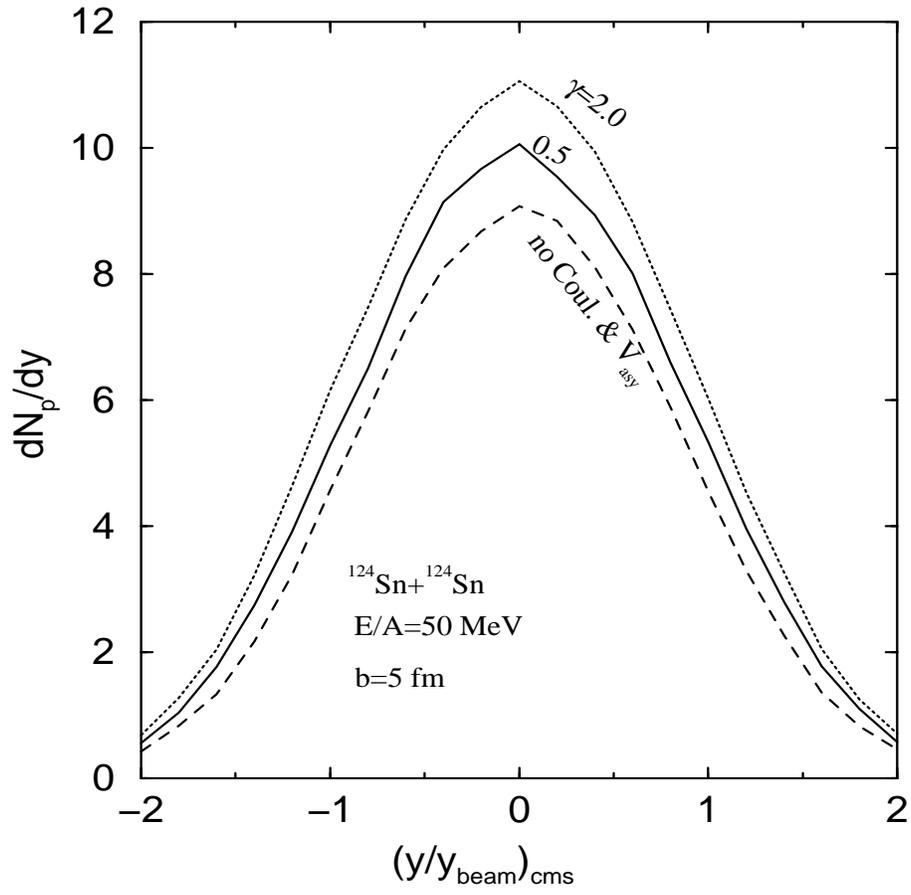,width=12cm,height=12cm,angle=-90} 
\caption{Rapidity distributions of unbound protons for the Sn+Sn reaction.}
\label{fig3}
\end{figure}

\begin{figure}[htp] 
\centering \epsfig{file=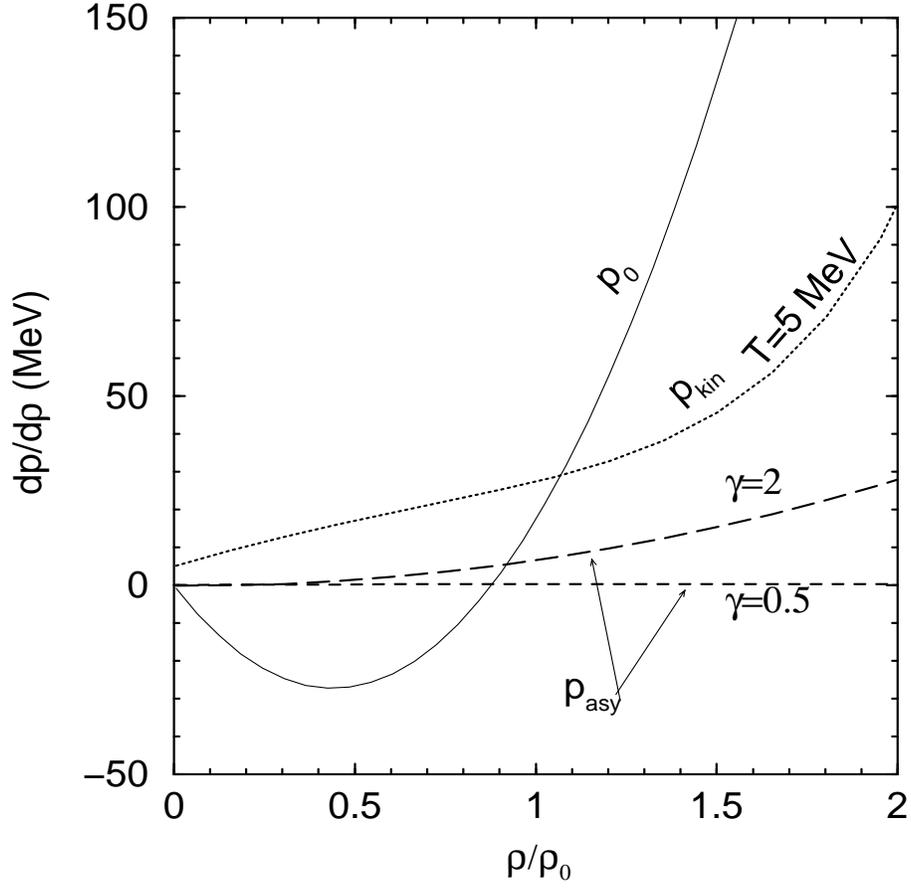,width=12cm,height=12cm,angle=-90} 
\caption{Derivatives of the three partial pressures as a function of density with 
the isospin asymmetry $\delta=0.2$ and the $\gamma$ parameter of 0.5 and 2. The
solid (dashed and long-dashed) line is for the isospin-independent (dependent) 
interaction pressure, while the dotted line is for the kinetic pressure at a temperature of
5 MeV.}
\label{fig4}
\end{figure} 

\begin{figure}[htp] 
\centering \epsfig{file=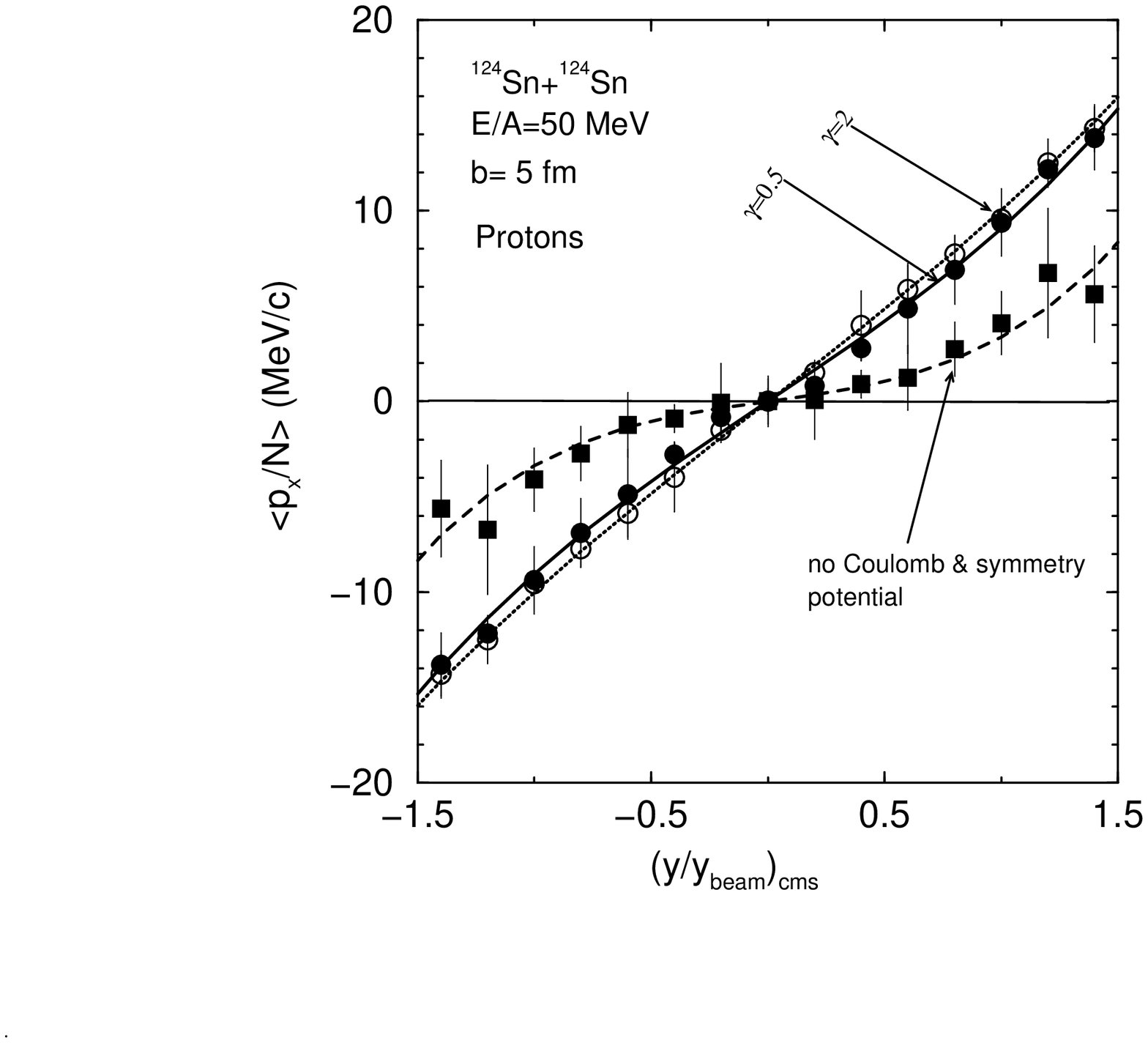,width=12cm,height=12cm,angle=-90} 
\caption{The average in-plane transverse momentum of free protons as a function 
of rapidity for the Sn+Sn reaction.}
\label{fig5}
\end{figure} 

\begin{figure}[htp] 
\centering \epsfig{file=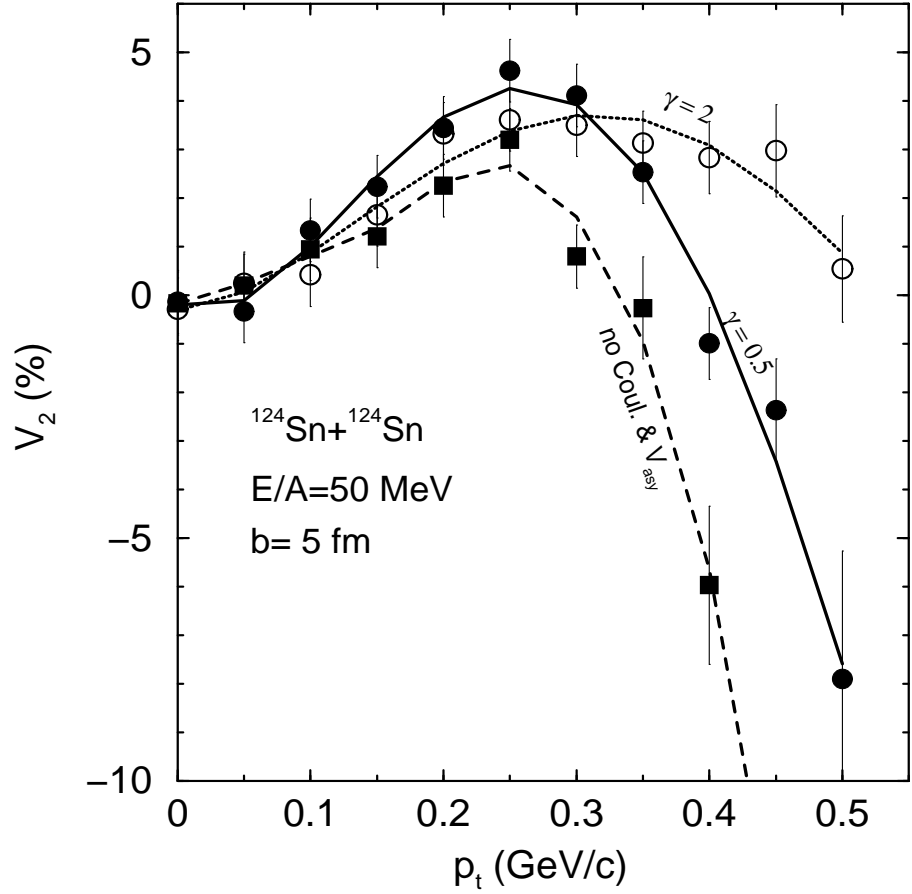,width=12cm,height=12cm,angle=-90} 
\caption{The elliptic flow for midrapidity free protons as a function of
transverse momentum for the Sn+Sn reaction.}
\label{fig6}
\end{figure}

\end{document}